\pdfoutput=1
\documentclass[runningheads]{./llncs}

\usepackage[latin1]{inputenc}
\usepackage{graphicx}
\usepackage{tabularx}
\usepackage[dvipsnames,usenames]{color}
\usepackage{listings}
\newcommand{\secend}{}

\title{%
  \textbf{%
    A Program Transformation for Continuation Call-Based Tabled
    Execution%
  } }

\titlerunning{%
    A Program Transformation for Continuation Call-Based Tabled
    Execution%
   }

\author{
  Pablo Chico de Guzman\inst{1} \ \
  \ \
  Manuel Carro\inst{1} \ \
  \ \
  Manuel V.\ Hermenegildo\inst{1,2}
  \ \ \\
\ \ 
  \email{pchico@clip.dia.fi.upm.es}
  \hspace{1ex}
  \email{\{mcarro,herme\}@fi.upm.es}
}

\authorrunning{P.\ Chico de Guzmán, M. Carro, M. Hermenegildo}

\institute{%
  School of Computer Science, Univ.\ Politecnica de Madrid,
  Spain
  \and
  IMDEA Software, Spain
}

\newcounter{mnotei}
\begin{document}

\setcounter{page}{75}

\maketitle

\begin{abstract}
  The advantages of tabled evaluation regarding program termination
  and reduction of complexity are well known ---as are the
  significant implementation, portability, and maintenance efforts
  that some proposals (especially those based on suspension)
  require. This implementation effort is reduced by program
  transformation-based continuation call techniques, at some
  efficiency cost.  However, the traditional formulation of this
  proposal by Ramesh and Cheng limits the interleaving of tabled and
  non-tabled predicates and thus cannot be used as-is for arbitrary
  programs.
  In this paper we present a complete translation for the continuation
  call technique which, using the runtime support needed for the
  traditional proposal, solves these problems and makes it possible to
  execute arbitrary tabled programs.  We present performance results
  which show that CCall offers a useful tradeoff that can be
  competitive with state-of-the-art implementations.

\end{abstract}

\section{Introduction}
\secend


Tabling~\cite{tamaki.iclp86,Warren92,chen96:tabled_evaluation} is a
strategy for executing logic programs which uses \emph{memoization} of
already processed calls and their answers to improve several of the
limitations of SLD resolution.
It brings termination for bounded term-size programs and improves
efficiency in programs which perform repeated computations and has
been successfully applied to deductive
databases~\cite{ramakrishnan93survey}, program
analysis~\cite{pracabsin,Dawson:pldi96-short}, reasoning in the
semantic Web~\cite{zou05:owl-tabling}, model
checking~\cite{ramakrishna97:model_checking_tabling}, etc.  

However, tabling also has certain drawbacks, including that predicates
to be tabled have to be selected carefully\footnote{XSB includes an
  \texttt{auto\_table} declaration which triggers a conservative
  analysis to detect which predicates are to be tabled in order to
  ensure termination.  However, more predicates than needed can be
  selected.}
in order not to incur in undesired slowdowns and, specially relevant to
our discussion, that its efficient implementation is generally
complex.  In \emph{suspension-based tabling} the computation state of
suspended tabled subgoals has to be preserved to avoid backtracking
over them.  This is done either by \emph{freezing} the stacks, as in
XSB~\cite{sagonas98:xsb-abstract-machine}, by copying to another area,
as in CAT~\cite{demoen98:cat}, or by using an intermediate solution as
in CHAT~\cite{demoen99:chat}.  \emph{Linear tabling} maintains instead
a single execution tree
without requiring suspension and resumption of sub-computations.  The
computation of the (local) fixpoint is performed by making subgoals
``loop'' in their alternatives until no more solutions are found.
This may make some computations to be repeated.
Examples of this method are the linear tabling of
B-Prolog~\cite{zhou:linear-tab-bprolog,zhou07:linear-tab-opt-tplp-prev-publ}
and the DRA scheme~\cite{guo01:DRA-iclp}.  Suspension-based mechanisms
achieve very good performance but, in general, require deeper changes
to the underlying 
implementation.  Linear mechanisms, on the other hand, can usually be
implemented on top of existing sequential engines without major
modifications.

The Continuation Call (\texttt{CCall}) approach to
tabling~\cite{DBLP:journals/tkde/RameshC97,rocha-iclp07} tries to
combine the best of both worlds: it is a reasonably efficient
suspension-based mechanism which requires relatively simple additions
to the Prolog implementation / compiler,\footnote{As an
  example, no modification to the underlying engine is needed.} thus making
maintenance and porting much easier.  In~\cite{improv-tabling-padl2008} we
proposed a number of optimizations to the \texttt{CCall} approach and
showed that with such optimizations performance could be competitive
with traditional implementations. However, this was only partially
satisfactory since the \texttt{CCall} tabling approach is restricted
to programs with a certain interleaving of tabled and
non-tabled predicate calls (see Figure~\ref{fig:tabled-environment2} and
Section~\ref{sec:worong-translation}), and thus cannot execute general  
tabled programs. 

In this paper
we present an extension of the \texttt{CCall} translation
which, using the same runtime support of the traditional proposal,
overcomes the problems pointed out above.
In Section~\ref{sec:comparing} we present a
complexity comparison of the proposed approach with CHAT. 
Finally, we present performance results from our implementation. These
results show that our approach offers a useful tradeoff 
which can be competitive with state of the art
implementations, while keeping implementation efforts relatively low.

\secend
\section{The Continuation Call Technique}
\label{sec:tabling-basics}
\secend

We sketch now how tabled
evaluation~\cite{chen96:tabled_evaluation,sagonas98:xsb-abstract-machine}
works from a user point of view and we briefly describe the
Continuation Call technique,
on which we base our work.

\secend
\subsection{Tabling Basics}
\label{sec:tabling-example}
\secend

We will use as example the program in
Figure~\ref{fig:tabled-environment}, whose purpose is to determine the
reachability of nodes in a graph.  If the graph contains cycles, there
will be queries which will make the program loop forever under the
standard SLD resolution strategy, regardless of the order of the
clauses.  Tabling changes the operational semantics for predicates
marked with the \emph{\texttt{:- table}} declaration, which forces the
compiler and runtime system to distinguish the first occurrence of a
tabled goal (the \emph{generator}) and subsequent calls which are
identical up to variable renaming (the \emph{consumers}).  The
generator applies resolution using the program clauses to derive
answers for the goal.  Consumers \emph{suspend} the current execution
path (using implementation-dependent means) and start execution on a
different branch corresponding to another clause of the predicate
within which the execution was suspended.  When such an alternative branch
finally succeeds, 
the answer generated for the initial query (the generator) is inserted in a table
associated with that generator. This makes it possible to
reactivate consumers and to continue execution at the point
where they were stopped. Thus, consumers do not use SLD resolution,
but obtain instead the answers from the table where they were
previously inserted by the generator.  Predicates not marked as tabled
are executed according to SLD resolution, hopefully with minimal
overhead due to the availability of tabling. This can be graphically
seen as the ability to suspend execution in a part of the tree which
cannot progress (because it enters a loop) and continue it somewhere
else, where a solution for the looping goal can be produced.

\secend
\subsection{\texttt{CCall} by Example}
\label{sec:original-ccall}
\secend

\texttt{CCall} implements tabling by a combination of program
transformation and side effects in the form of insertions into and
retrievals from a table which relates calls, answers, and the
continuation code to be executed after consumers read answers from the
table.  We will now sketch how the mechanism works using the
\texttt{path/2} example (Figure~\ref{fig:tabled-environment}). The
original code is transformed into the program in
Figure~\ref{fig:tabled-environment-transformed} which is the one
actually executed.

Roughly speaking, the transformation for tabling is as follows: an
auxiliary predicate (\texttt{slg\_path/2}) for \texttt{path/2} is
introduced so that calls to 
\texttt{path/2} made from regular (SLD) Prolog execution do not need
to be aware of the fact that \texttt{path/2} is being tabled.  The
primitive \texttt{slg/1} will make sure that its argument is executed
to completion and will return, on backtracking, all the solutions
found for the tabled predicate.  To this end, \texttt{slg/1} checks if
the call has already been executed. If so, all its answers are
returned by backtracking. Otherwise, control is passed to a new
predicate (\texttt{slg\_path/2} in this case).\footnote{The unique name
  has been created for simplicity by prepending \texttt{slg\_} to the
  predicate name --any safe means of constructing a unique predicate
  symbol 
  can be used.}  \texttt{slg\_path/2} receives in its first argument
the original call to \texttt{path/2} and in the second argument the
identifier of its generator, which is used to relate operations on the
table with this initial call.  Each clause of \texttt{slg\_path/2} is
derived from a clause of the original \texttt{path/2} predicate by:

\secend
\begin{itemize}
\item Adding an \texttt{answer/2} primitive at the end of each clause
  of the original tabled predicate. \texttt{answer/2} is responsible
  for inserting answers in the table after checking for redundancy.
\item Instrumenting calls to tabled predicates using the
  \texttt{slgcall/1} primitive.  If this tabled call is a consumer,
  \texttt{path\_cont/3}, along with its arguments,
  is recorded as (one of) the continuation(s) of 
  its generator. If the tabled call is a generator, it is associated
  with a new call identifier and execution follows using the
  \texttt{slg\_path/2} program clauses to derive new answers (as done
  by \texttt{slg/1}).  Besides, \texttt{path\_cont/3} will be recorded
  as a continuation of the generator identified by \texttt{Id} if the
  tabled call cannot be completed (there were dependencies on previous
  generators).  The \texttt{path\_cont/3} continuation will be called
  consuming found answers or erased upon completion of its generator.
\item Encoding the remaining of the clause body of \texttt{path/2}
  after the recursive call by using \texttt{path\_cont/3}. It is
  constructed  similarly to \texttt{slg\_path/2}, i.e., applying
  the same transformation as for the initial clauses and calling
  \texttt{slgcall/1}.
\end{itemize}

\begin{figure}[t]
  \centering
  \begin{minipage}[b]{0.36\linewidth}
    \begin{lstlisting}
@\neck@ table path/2.

path(X, Z)@\neck@ 
    edge(X, Y), 
    path(Y, Z).
path(X, Z)@\neck@
    edge(X, Z).
    \end{lstlisting}
  \vspace{5mm}
  \caption{A sample program.}
  \label{fig:tabled-environment}
  \end{minipage}
\hfill
  \begin{minipage}[b]{0.6\linewidth}
    \begin{lstlisting}
path(X, Y)@\neck@ slg(path(X, Y)).
slg_path(path(X, Y), Id)@\neck@
    edge(X, Y),
    slgcall(path_cont(Id, [X], path(Y, Z))).
slg_path(path(X, Y), Id)@\neck@
    edge(X, Y),
    answer(Id, path(X, Y)).

path_cont(Id, [X], path(Y, Z))@\neck@
    answer(Id, path(X, Z)).
    \end{lstlisting}
    \caption{The program in Figure~\ref{fig:tabled-environment} after
      being transformed for tabled execution.} 
  \label{fig:tabled-environment-transformed}
  \end{minipage}
\end{figure}

\secend
The second argument of \texttt{path\_cont/3} is a list of bindings
needed to recover the environment of the continuation call.  Note
that, in the program in Figure~\ref{fig:tabled-environment}, an answer
to a query such as \texttt{?- path(X, Y)} may need to bind variable
\texttt{X}.  This variable does not appear in the recursive call to
\texttt{path/2}, and hence it does not appear in the \texttt{path/2}
term passed on to \texttt{slgcall/1} either.  In order for the body of
\texttt{path\_cont/3} to insert in the table the answer corresponding
to the initial query, variable \texttt{X} (and, in general, any other
necessary variable) has to be passed down to \texttt{answer/2}.  This
is done with the list \texttt{[X]}, which is inserted in the table as
well and completes the environment needed for the continuation
\texttt{path\_cont/3} to resume the previously suspended call.

A safe approximation of the variables which should appear in this list
is the set of variables which appear in the clause before the tabled
goal and which are used in the continuation, including the
\texttt{answer/2} primitive.  Variables appearing in the tabled call
itself do not need to be included, as they will be passed along
anyway. This list of bindings corresponds to the frame of the parent
call if the \texttt{answer/2} primitive is added to the end of the
body being translated. More details about \texttt{CCall} approach and
their primitives can be found at~\cite{DBLP:journals/tkde/RameshC97}.

\secend
\secend
\subsubsection{Key Contribution of CCall:}
a new predicate name is created for all points where suspension can
happen.  Suspension is performed by saving this predicate name, a list
of bindings, and a generator identifier.  Resumption is performed by
constructing a Prolog goal with the information
saved on suspension plus the answer which raised the resumption.  It
is clear that this is 
significantly simpler to implement than other approaches as XSB or
CHAT, where changes in the abstract machine have to be introduced.
Consequently, porting and maintainability are simpler too,
since \texttt{CCall} is independent of the compiler and how to create
a Prolog term on the heap is the only one low level operation to
implement.

\secend
\section{Mixing Tabled and Non-Tabled Predicates}
\label{sec:complementing-translation}
\secend

A continuation is the way \texttt{CCall} tabling preserves both the
environment and the code of a consumer to be resumed.  The list of
bindings contains the same variables as the frame of the predicate
where the \texttt{slgcall/1} primitive is executed, taking into
account the \texttt{answer/2} primitive added at the end of the
clause.  However, the \texttt{CCall} approach to tabling, as
originally proposed, has a problem when Prolog predicates appear
between generators and consumers: the environments created by the
non-tabled predicates are not taken into account, and they may be
needed to correctly suspend and resume tabled predicates, as the
example in the following section shows.


\secend
\secend
\secend
\subsection{An Ill-Behaved Transformation}
\label{sec:worong-translation}
\secend
\secend

Figure~\ref{fig:tabled-environment2} shows an example of a tabled
program, where tabled and non-tabled execution (\texttt{t/1} and
\texttt{p/1}) are mixed. The translation of the program is shown in
Figure~\ref{fig:tabled-environment-transformed2}, taking into account
the rules in Section~\ref{sec:original-ccall}.

\begin{figure}[t]
  \centering
  \begin{minipage}[b]{0.45\linewidth}
    \begin{lstlisting}
@\neck@ table t/1.

t(A)@\neck@ 
    p(B), 
    A is B + 1.
t(0).

p(B)@\neck@ t(B), B < 1.
    \end{lstlisting}
  \caption{A program for which the original \texttt{CCall}
    transformation fails.}
  \label{fig:tabled-environment2}
  \end{minipage}
\hfill
  \begin{minipage}[b]{0.5\linewidth}
    \begin{lstlisting}
t(A)@\neck@ slg(t(A)).
slg_t(t(A), Id)@\neck@
    p(B), A is B + 1,
    answer(Id, t(A)).

slg_t(t(0), Id)@\neck@
    answer(Id, t(0)).

p(B)@\neck@ t(B), B < 1.
    \end{lstlisting}
  \caption{The program in Figure~\ref{fig:tabled-environment2} after
    being transformed for tabled execution.}
  \label{fig:tabled-environment-transformed2}
  \end{minipage}
\end{figure}

\begin{figure}[t]
  \centering
    \fbox{
    \raisebox{-2ex}{\parbox{0.8\linewidth}{%
        \centering\includegraphics[scale=0.55]{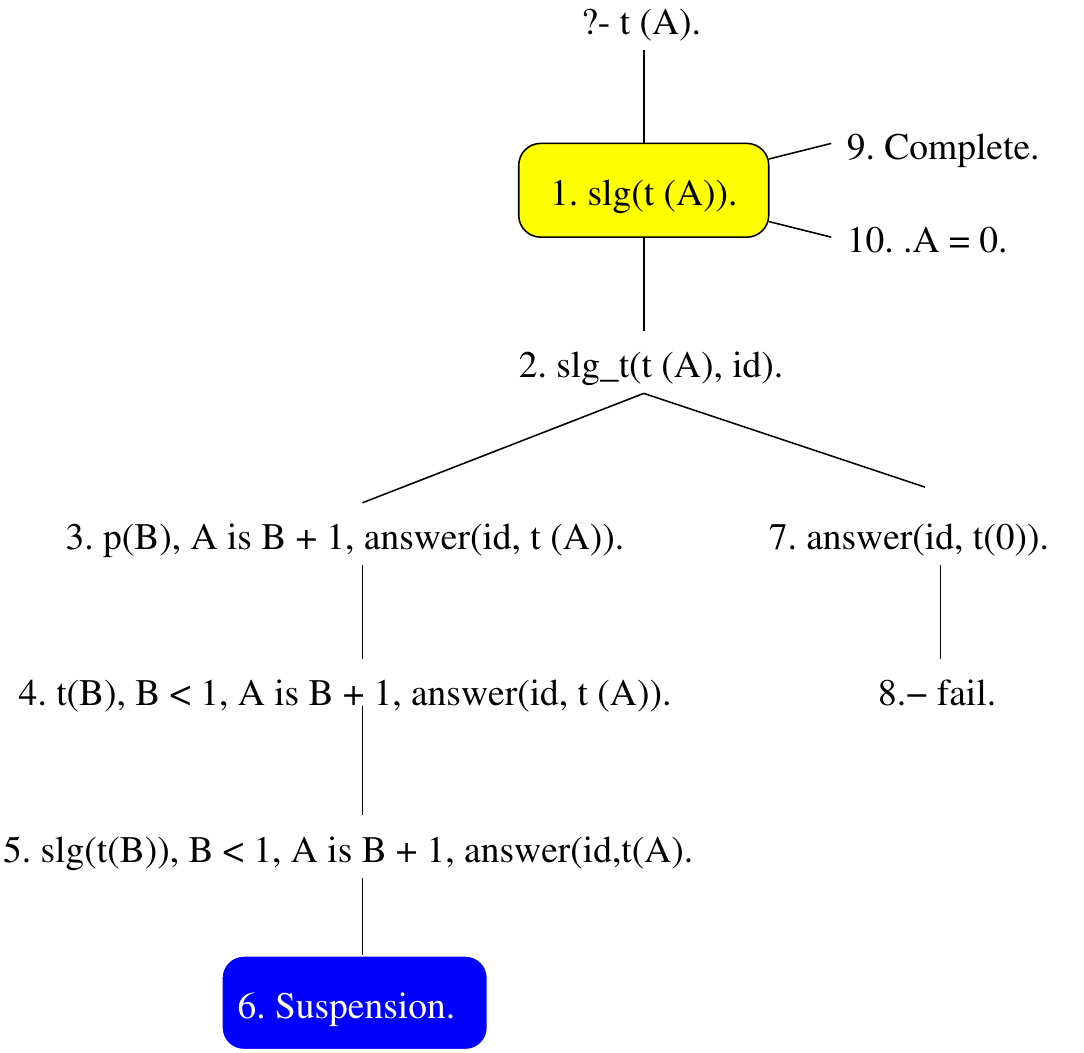}}}}
  \caption{Tabling execution of example of
    Figure~\ref{fig:tabled-environment}.}
  \label{fig:wrong-translation}
\end{figure}

The execution of the program with the query \texttt{t(A)} is shown in
Figure~\ref{fig:wrong-translation}. The execution is correct until
\texttt{slg/1} is called again by \texttt{p/1}.  At that point
execution should suspend (and later resume), but \texttt{slg/1}
does not have any associated continuation, and it does not have any
pointer to the code to be executed on resumption (partially in
\texttt{p/1} and partially in \texttt{slg\_t/2}): \texttt{B < 1, A is
  B + 1, answer(Id,t(A))} is lost on backtracking and it is not
reachable when resuming. Consequently, the second answer to the query,
\texttt{t(1)}, is lost.


The call to \texttt{t(B)} made by \texttt{p(B)} could have been translated as
if it were in the body of a tabled clause, but in that case the piece of code
\texttt{A is B + 1} in the first clause of \texttt{t/1} would be lost anyway.
This is an
example of why all the frames between a consumer and its nearest
generator have to be saved when suspending, and it is not enough to
save just the last one, as in the original \texttt{CCall}
proposal~\cite{DBLP:journals/tkde/RameshC97}, which
does work, however,
when all the calls to the tabled predicates appear in the body of the
clause of a tabled predicate.  In that case, it is enough to save the
last frame with the associated continuation code.  Note that all the
suspension-based tabling approaches preserve the frames / environments from the
consumer until the corresponding generator.

To solve this problem, we have extended the translation to take into
account a new kind of predicates, named \emph{bridges}. A
\textrm{bridge} predicate is a non-tabled Prolog 
predicate whose clauses generate frames which have to be saved in the
continuation of a consumer. In the example of
Figure~\ref{fig:tabled-environment2}, \texttt{p/1} is a bridge
predicate.


\subsection{Marking Predicates as Bridges}
\label{sec:which-bridges}
\secend



Bridge predicates are all the non-tabled predicates which can appear
in the execution tree of a query between a generator and each of its
consumers, i.e., the predicates whose environments are in the
local stack between the environment of the generator and the environment
of each of its consumers.  Note that tabled predicates do not need to
be included as bridge predicates as their environment will be already
saved by the translation.  Additionally, only recursive calls which
can lead to infinite loops under SLD resolution have to actually be
taken into account, because these are the only ones which can suspend
and later be resumed.  Programs for which tabling merely speeds up
already terminating computations are not subject to the problem
outlined above, and therefore do not benefit from the improved
translation shown herein.


Thus, in order to determine a minimal set of bridge predicates,
\texttt{$B_{min}$}, we need to determine before the minimum set of
tabled predicates, \texttt{$T_{min}$}, which ensures termination.
When \texttt{$T_{min}$} is found, \texttt{$B_{min}$} is the set of
non-tabled predicates which are ``in the middle'' of two calls to
predicates belonging to \texttt{$T_{min}$}.
Since looking for \texttt{$T_{min}$} is undecidable (because it
implies detecting infinite failures), looking for \texttt{$B_{min}$}
is also undecidable and a \emph{safe approximation}, which may
mark as bridge some predicates which do not need to be, is
needed. 
%

As we will see in Section~\ref{sec:new-translation}, the only disadvantage
of such an over-approximation is that some code will be duplicated (to
accept a new argument for the case where a bridge predicate is called
from a tabled execution), and that bridge predicates, having an extra
argument, can be called when this is not needed.  The algorithm we
have implemented (Figure~\ref{fig:bridges-code}) only looks for tabled
predicates which can recursively call themselves.  For the examples
used for performance evaluation in Section~\ref{sec:perf-eval}, using
the safe approximation algorithm produces an average slowdown of only
3\% with respect to a perfect characterization of bridge predicates.


\begin{figure}[t]
  \centering
    \begin{lstlisting}
  Make a graph $G$ with an edge ($p1/n1$, $p2/n2$) $\Leftrightarrow$ $p2/n2$ is called from $p1/n1$
  $Bridges$ = $\emptyset$
  FOR each predicate $T$ in TABLED PREDICATES
    $Forward$ = All predicates reached from $T$ in $G$
    $Backward$ = All predicates from which $T$ is reached in $G$
    $Bridges$ = $Bridges$ $\cup$ ($Forward$ $\cap$ $Backward$)
  $Bridges$ = $Bridges$ - TABLED PREDICATES
  \end{lstlisting}
  \caption{Safe approximation to look for bridge predicates.}
  \label{fig:bridges-code}
\end{figure}

\secend
\section{A General Translation for Tabled Programs}
\label{sec:new-translation}
\secend

In this section we present program transformation rules which take
into account bridge predicates. This transformation assumes that the
safe approximation algorithm for bridge predicates has already been
run, and all the bridge predicates have been marked by adding a
\texttt{:-~bridge~P/N} declaration in the program.

As seen in Section~\ref{sec:original-ccall}, a continuation is the way
to save an environment, because the predicate name is the same as the
PC counter of the environment and the list of bindings is the same as
the variables that a environment saves. Consequently, the goal of the
new translation is to associate a continuation with each of the bridge
predicates to save their associated environment. These continuations receive a 
new argument (the continuation to be executed) which is used to push a
pointer (i.e., the name of a predicate) to the code to continue with,
in a way similar to environments in local stacks.


\secend
\secend
\subsection{Translation Rules}
\label{sec:translation-rules}
\secend

The rules for the original translation have three different goals: to
maintain the interface with the rest of the code, to manage tabled
calls which appear in the body of the clauses of a tabled predicate,
and to insert answers at the end of the evaluation of each clause. The
same points have to be addressed for bridge clauses, taking into
account that a tabled or bridge call has to be translated if it
appears in the body of a tabled predicate or a bridge predicate.

\begin{figure}[t]
  \centering
    \begin{lstlisting}
trans(C, C) :- \+ table(C), \+ bridge(C).  
trans(( :- table P/N ), ( P(X1..Xn) :- slg(P(X1..Xn)) )).
trans(( Head :- Body ), LC) :- 
   table(Head), 
   Head_tr =.. ['slg_' $\circ$ Head, Head, Id],
   End = answer(Id, Head),
   transBody(Head_tr, Body, Id, [], End, LC). 
trans(( Head :- Body ), ( Head :- Body ) $\circ$ LC) :- 
   bridge(Head),
   Head_tr =.. [Head $\circ$ '_bridge', Head, Id, Cont],
   End = call(Cont),         
   transBody(Head_tr, Body, Id, Cont, End, LC). 

transBody([], [], _, _, [], []).
transBody(Head, Body, Id, ContPrev, End, ( Head :- Body_tr ) $\circ$ RestBody_tr) :-
   following(Body, Pref, Pred, Suff),
   getLBinds(Pref, Suff, LBinds),
   updateBody(Pred, End, Id, Pref, LBinds, ContPrev, Cont, Body_tr),
   transBody(Cont, Suff, Id, ContPrev, End, RestBody_tr).

following(Body, Pref, Pred, Suff) :-
   member(Body, Pred),
   (table(Pred); bridge(Pred)), !, 
   Body = Pref $\circ$ Pred $\circ$ Suff.

updateBody([], End, _Id, Pref, _LBinds, _ContPrev, [], Pref $\circ$ End).
updateBody(Pred, _End, Id, Pref, LBinds, ContPrev, Cont, Pref $\circ$ slgcall(Cont)) :-
   table(Pred),
   getNameCont(NameCont),
   Cont = NameCont(Id, LBinds, Pred, ContPrev).
updateBody(Pred, _End, Id, Pref, LBinds, ContPrev, Cont, Pref $\circ$ Bridge_call) :-
   bridge(Pred),
   getNameCont(NameCont),
   Cont = NameCont(Id, LBinds, Pred, ContPrev),
   Bridge_call =.. [Pred $\circ$ '_bridge', Pred, Id, Cont] .
    \end{lstlisting}
  \caption{The Prolog code of the translation rules.}
  \label{fig:trans-code}
\end{figure}

The rules for the new translation, which uses the same primitives as
the original 
\texttt{CCall} proposal, are shown in Figure~\ref{fig:trans-code},
where for conciseness we have used a sugared Prolog-like language.
For example, a functional
syntax is implicitly assumed where needed, and infix '$\circ$' is a
general \texttt{append} function which joins either (linear)
structures or, when applied to atoms, concatenates them.  It may
appear in an output head position with the expected semantics.

The \texttt{trans/2} predicate receives a clause to be translated and
returns the list of clauses resulting from the translation. Its first
clause ensures that predicates which are non-tabled and non-bridge are
not transformed.\footnote{The predicates \texttt{table/1} and
  \texttt{bridge/1} are dynamically generated by the compiler from the
  corresponding declaration.  They check if their argument is a clause
  of a tabled or bridge predicate, or if their argument is a functor
  corresponding to a tabled or bridge predicate, respectively.} The
second one is to generate the interface of table predicates with the
rest of the code: if there is a tabled declaration, the interface is
generated. The third clause translates clauses of tabled predicates,
and the fourth one translates clauses of bridge predicates, where the
original one is maintained in case it is called outside a tabled call
(this is in order to preserve the interface with non-tabled code).
They generate the new head of the clause, \texttt{Head\_tr}, and the
code which has to be appended at the end of the body, \texttt{End},
before calling \texttt{transBody/6} with these arguments. \texttt{End}
can be the \texttt{answers/2} primitive for tabled clauses or
\texttt{call(Cont)}, which invokes the following pushed continuation,
stored in the fourth argument.

\texttt{transBody/6} generates, in its last argument, the translation
of the body of a clause by taking care, in each iteration, of the code
until the next tabled or bridge call, or until the end the clause, and
appending the translation of the rest of the clause to this partial
translation. In other words, it calls \texttt{updateBody/8} to
translate tabled or bridge calls and continues translating the rest of
the body.

The \texttt{following/4} splits a clause body in three parts: a
prefix, until the first time a tabled or bridge call appears, the
tabled or bridge call itself, and a suffix from this call until the
end of the clause.  \texttt{getLBinds/3} obtains the list of variables
which have to be saved to recover the environment of the consumer,
based on the ideas of Section~\ref{sec:original-ccall}.

The \texttt{updateBody/8} predicate completes the body prefix until
the next tabled or bridge call. Its first six arguments are inputs,
the seventh one is the head of the continuation for the suffix of the
body, and the last argument is the new translation for the prefix.
The first clause takes care of the base case, when there are no calls
to bridge or tabled predicates left,
the second clause generates code for a call to a
tabled predicate, and the last one does the same with a bridge
predicate.  That \texttt{getNameCont/1} generates a unique name for the
continuation.

We will now use the example in Figure~\ref{fig:tabled-environment2},
adding a \texttt{:- bridge p/1} declaration, to exemplify how a
translation would take place.

\secend
\subsection{The Previous Example with the Correct Transformation}
\label{sec:correct-translation}
\secend


The translation of the first clause of \texttt{t/1} is done by the
third clause of \texttt{trans/2}, which makes the head of the
translated clause to be \texttt{slg\_t(t(A), Id)} and states that the
final call of that clause has to be \texttt{answer(Id, t(A))} ---i.e.,
when the clause successfully finishes, it adds the answer to the
table.

\texttt{transBody/6} takes care then of the rest of the body, which
identifies which environment variables (\texttt{A}, in this case) have
to be saved and matches \texttt{Pref}, \texttt{Pred}, and
\texttt{Suff} with the goals before the call to the bridge predicate
(none --- and empty conjunction), the call to the bridge predicate
(\texttt{p(B)}), and the goals after this call (\texttt{A is B + 1}).
The third clause of \texttt{updateBody/8} generates the body of
\texttt{Head\_tr}, to give the first clause of
\texttt{slg\_t/2}.  A continuation is generated for the rest of
the body; the code of the continuation is a predicate whose head is
\texttt{slg\_t0/3} and its body is generated by 
the first clause of \texttt{updateBody/8}.

The translation of the second clause of \texttt{t/1} is simpler, as it
only has to add \texttt{answer(Id, t(0))} at the end of the body of
the new predicate.

\begin{figure}[t]
  \centering
  \begin{minipage}[b]{0.45\linewidth}
    \begin{lstlisting}
t(A) :- slg(t(A)).
slg_t(t(A), Id) :-
   p_bridge(p(B), Id, slg_t0(Id, [A], p(B), [])).

slg_t(t(0),Id) :- answer(Id, t(0)).

slg_t0(Id, [A], p(B), []) :- 
   A is B + 1,
   answer(Id, t(A)).
    \end{lstlisting}
  \end{minipage}
\hfill
  \begin{minipage}[b]{0.45\linewidth}
    \begin{lstlisting}
p(B) :- t(B), B < 1.

p_bridge(p(B), Id, Cont) :- 
   slgcall(p_bridge0(Id, [], t(B), Cont)).

p_bridge0(Id, [], t(B), Cont) :-
   B < 1,
   call(Cont).
    \end{lstlisting}
  \end{minipage}
  \caption{The program in Figure~\ref{fig:tabled-environment2} after
    being transformed for tabled execution.}
  \label{fig:tabled-environment-transformed3}
\end{figure}



The clause for \texttt{p/1} is kept to maintain its interface when it
is not called from inside a another tabled execution.  The translation
for the clause of \texttt{p/1} is made by the fourth clause of
\texttt{trans/2} where \texttt{Head\_tr} is unified with
\texttt{p\_bridge(p(B), Id, Cont)}.  \texttt{End} is unified with
\texttt{call(Cont)} --- a call to the continuation code to be resumed by
the following pushed continuation.
\texttt{transBody/6} finds an empty list of environment variables and
unifies \texttt{Pref}, \texttt{Pred} and \texttt{Suff} with
\texttt{[]}, \texttt{t(B)} and \texttt{B < 1}, respectively.  The
second clause of \texttt{updateBody/8} generates the body for the new
predicate
\texttt{p\_bridge/3}.  A continuation is
generated to execute the rest of the body, whose head is
\texttt{p\_bridge0/3} and whose body is generated by the first clause of
\texttt{updateBody/8}.
As we can see, bridge predicates are pushing continuations which are
sequentially called when consumers are resumed.

\secend
\subsection{Execution of the Transformed Program}
\label{sec:new-execution}
\secend

\begin{figure}[t]
  \centering
    \fbox{
    \raisebox{-2ex}{\parbox{\linewidth}{%
        \centering\includegraphics[scale=0.50]{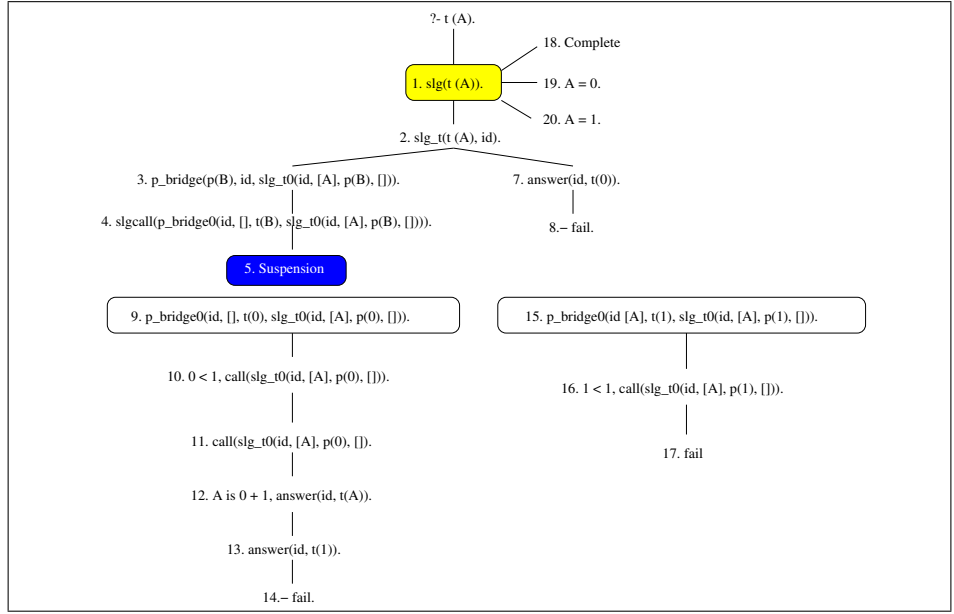}}}}
  \caption{New \texttt{CCall} tabling execution.}
  \label{fig:bridge-execution}
\end{figure}

The execution tree of the transformed program is shown in
Figure~\ref{fig:bridge-execution}. It is similar to that in
Figure~\ref{fig:wrong-translation}, but a continuation
\texttt{slg\_t0(id, [A], p(B), [])} is passed to the transformed
clause of \texttt{p/1}. This continuation contains the code to be
executed after the execution of \texttt{p(B)} and the list
\texttt{[A]} needed to recover its environment. Consequently, there
are two continuations associated with the suspension: one continuation
to execute the rest of the code of \texttt{p(B)} and another one to
execute the rest of the code of \texttt{t(A)}.

After the first answer is found, this double continuation is
resumed. It is executed as a normal Prolog and the second
answer, \texttt{t(1)}, is found. 

\secend
\section{$\Theta$(CHAT) is not comparable with $\Theta$(CCall)}
\label{sec:comparing}
\secend

In this section we present a comparative analysis of the complexity of
\texttt{CCall} and CHAT, which is an efficient implementation
of tabling with a comparatively simple machinery.  Since it is known
that $\Theta$(CHAT) is
$\Theta$(\texttt{SLG-WAM})~\cite{demoen99:chat_complexity}, the
comparative analysis applies to the \texttt{SLG-WAM} as well.

The complexity analysis focuses on the operations of suspension and
resumption.  The environment of a consumer has to be protected when
suspending to reinstall it when resuming. \texttt{CCall} achieves that
by copying the continuation associated with the consumer in a special
memory area to be protected on backtracking.  In the original
implementation~\cite{DBLP:journals/tkde/RameshC97} this continuation
is copied from the heap to a separate table (when suspending) and back
(when resuming).  As proposed in~\cite{improv-tabling-padl2008},
continuations can be saved in a special memory area with the same data
format as the heap.  This makes it possible to use WAM instructions
and additional machinery on them and, when resuming, they can be used
as normal Prolog data and code, without being recopied each time a
consumer is resumed.

On the other hand, CHAT freezes the heap and the frame stack
when resuming. The heap and frame stack are frozen by traversing the
choice point stack. For all the choice points between the consumer
choice point and its generator, the pointer to the end of the heap and
frame stack are changed to the values of the consumer choice point
values. By doing that, heap and frame stack are protected on
backtracking.  However, the consumer choice point has to be copied to
a special memory area as well as the segment trail (with its
associated values) between the consumer and the generator, to
reinstall the values of the bound variables at the time of suspension
which backtracking will unbind. In consequence, when resuming the
trail values have to be reinstalled as well as the consumer choice
point.


Each consumer is suspended only once, and it can be resumed several
times. The rest of the operations, i.e., checking if a tabled call is
a generator or a consumer, are not analyzed, because they are common
to both systems. In addition, we will ignore the cost of working at
the Prolog level, since this is an orthogonal issue: \texttt{CCall}
primitives could be compiled to WAM instructions and working at Prolog
level does not increase the system complexity.

\secend
\secend
\subsubsection{$\Theta$(\texttt{CCall}):}
\label{sec:o-ccall}
\secend

when suspending, \texttt{CCall} has to copy all the environments until the
last generator and the structures in the heap which hang from them.
If we name \texttt{E} the size of all the environments and \texttt{H} the
size of the structures in the heap, the time consumption when
suspending is: \texttt{$\Theta$(E + H)}.

When resuming, \texttt{CCall} just has to perform pattern matching of
the continuation against its clause. The time taken by the pattern
matching depends on the size of the list of bindings, which is known
to be \texttt{$\Theta$(E)}. Since each consumer can be resumed
\texttt{N} times, the time consumption of resuming consumers is
\texttt{$\Theta$(N$\times$E)}.

\secend
\secend
\subsubsection{$\Theta$(CHAT):}
\label{sec:o-chat}
\secend

when suspending, CHAT has to traverse the frame and
choicepoint stacks, but with the improvements presented in
\cite{demoen99:chat_complexity}, the time this takes can be neglected
because a choice point is only traversed once for all the consumers.
The trail and the last choice point have to be copied. If we call
\texttt{T} the size of the trail and \texttt{C} the size of the choice
point, which is bound by a constant for a given program, the time
consumption when suspending is: \texttt{$\Theta$(T)}.

When resuming, CHAT has to reinstall the values of the frame
and the choice point. Since each consumer can be resumed \texttt{N}
times, the time consumption of resuming is \texttt{$\Theta$(N$\times$T)}.

\secend
\secend
\subsubsection{Analyzing the worst cases of both systems:}
\label{sec:o-worst-case}
\secend

we can conclude \texttt{E + H $\geq$ T}, because each variable can
only be once in the trail, and then \texttt{CCall} is worse than
CHAT when suspending. On the other hand, in case that
\texttt{E < T}, \texttt{CCall} is better than CHAT when
resuming. Consequently, for a plausible general case, the more
resumptions there are, the better \texttt{CCall} behaves in comparison
with CHAT, and conversely.
In any case, the worst and best cases for each implementation are
different, which makes them difficult to compare. For example, if
there is a very large structure pointed to from the environments, and
none of its elements are pointed to from the trail, \texttt{CCall} is
slower than CHAT, since it has to copy all the structure in a
different memory area when suspending and CHAT does nothing
both when suspending and when resuming.

On the other hand, if all the elements of the structure are pointed to
from the trail, \texttt{CCall} has to copy all the structure on
suspension in a different memory area to protect it on backtracking,
but it is ready to be resumed without any other operation (just a
unification with the pointer to the structure). CHAT has to copy all
the structure on suspension too, because all the structure is in the
trail. In addition, each time the consumer is resumed, all the
elements of the structure have to be reinstalled using the trail, and
CHAT has to perform more operations than \texttt{CCall}, and then, the
more resumptions there are, the worse CHAT would be in comparison with
\texttt{CCall}. Anyway, as the trail is usually much smaller than the
heap, in general cases, CHAT will have an advantage over
\texttt{CCall}.

\section{Performance Evaluation}
\label{sec:perf-eval}
\secend

We have implemented the proposed technique as an extension of the Ciao
system~\cite{ciao-reference-manual-1.13-short}. Tabled evaluation is
provided to the user as a loadable \emph{package} that implements the
new directives and user-level predicates, performs the program
transformations, and links in the low-level support for tabling.  We
have implemented \texttt{CCall} tabling with the efficiency
improvements presented in~\cite{improv-tabling-padl2008} and the new
translation for general programs explained in this paper.

Table~\ref{tab:comparison-xsb} aims at determining how the proposed
implementation of tabling compares with state-of-the-art systems
---namely, the latest available versions of XSB, YapTab, and B-Prolog,
at the time of writing, using the typical benchmarks which appear in
other performance evaluations
of tabling approaches.\footnote{This is in contrast
  to~\cite{improv-tabling-padl2008} where, due to the limitations  of
  the \texttt{CCall} approach the 
  benchmarks presented did not need the use of bridge predicates.}
In this table we provide, for several benchmarks, the raw time (in
milliseconds) taken to execute them using tabling.  Measurements have
been made with Ciao-1.13, using the standard, unoptimized
bytecode-based compilation, and with the \texttt{CCall} extensions
loaded, as well as in XSB 3.0.1, YapTab 5.1.1, and B-Prolog 7.0.  Note
that we did not compare with CHAT, which was available as a
configuration option in the XSB system and which was removed in recent
XSB versions.  CHAT can be expected to be at least as fast (if not
slightly faster) than XSB.

All the executions were performed using local scheduling and disabling
garbage collection; in the end this did not impact execution times
very much.  We used \texttt{gcc 4.1.1} to compile all the systems, and
we executed them on a machine with Fedora Core Linux, kernel 2.6.9,
and an Intel Xeon DESCHUTES processor.


\begin{table}[t]
  \centering
  \begin{tabular}{|l|r|r|r|r|c|} 
    \hline
    Program  & CCall  & XSB      & YapTab   & BProlog & \# Bridges
    \\\hline\hline                                 
    path     & 517.92 & 231.4  & 151.12 & 206.26 & 0
    \\\hline                                                              
    tcl      & 96.93  & 59.91  & 39.16  & 51.60  & 0
    \\\hline                                                             
    tcr      & 315.44 & 106.91 & 90.13  & 96.21  & 0
    \\\hline                                                             
    tcn      & 485.77 & 123.21 & 85.87  & 117.70 & 0
    \\\hline                                                                
    sgm      & 3151.8 & 1733.1 & 1110.1 & 1474.0 & 0
    \\\hline                                                                
    atr2     & 689.86 & 602.03 & 262.44 & 320.07 & 0
    \\\hline                                                                
    pg       & 15.240 & 13.435 & 8.5482 & 36.448 & 6
    \\\hline                                                                
    kalah    & 23.152 & 19.187 & 13.156 & 28.333 & 20
    \\\hline
    gabriel  & 23.500 & 19.633 & 12.384 & 40.753 & 12
    \\\hline
    disj     & 18.095 & 15.762 & 9.2131 & 29.095 & 15
    \\\hline
    cs\_o    & 34.176 & 27.644 & 18.169 & 85.719 & 14
    \\\hline
    cs\_r    & 66.699 & 55.087 & 34.873 & 170.25 & 15
    \\\hline
    peep     & 68.757 & 58.161 & 37.124 & 150.14 & 10
    \\\hline                                                             
\end{tabular}
\vspace{5mm}
\caption{Comparing Ciao+CCall with XSB, YapTab, and B-Prolog.}
\label{tab:comparison-xsb}
\end{table}

The first benchmark is \texttt{path}, the same as
Figure~\ref{fig:tabled-environment}, which has been executed with a
chain-shaped graph. Since this is a tabling-intensive program with no
consumers in its execution, the difference with other systems is
mainly due to having large parts of the execution done at Prolog
level. The following five benchmarks, until \texttt{atr2}, are also
tabling intensive. As their associated environments are very small,
\texttt{CCall} is far from its worst case (see
Section~\ref{sec:comparing}), and the difference with other systems is
similar to that in \texttt{path} and for a similar reason. The worst
case in this set is \texttt{tcn} because there are two calls to
\texttt{slgcall/1} per generator, and the overhead of working at the 
Prolog level is duplicated.

B-Prolog, which uses a linear tabling approach, suffers if costly
predicates have to be recomputed: this is what happens in benchmarks
from \texttt{pg} until \texttt{peep}, where tabled and non-tabled
execution is mixed.  This is a well-known disadvantage of linear
tabling techniques which does not affect suspension-based approaches.
It has to be noted, however, that latest versions of B-Prolog
implement an optimized variant of its original linear tabling
mechanism~\cite{zhou07:linear-tab-opt-tplp-prev-publ} which tries to
avoid reevaluation of looping subgoals.

In order to compare our implementation with XSB and YapTab, we must
take into account that the speeds of  XSB, and
YapTab\footnote{Note that we are comparing the tabled-enabled version
  of Yap, which is somewhat slower than the regular Yap.} are different,
at least in those cases where the program execution is large enough to
be really significant  (between 1.8 and 2 times slower in the case of
XSB and 1.5 times faster in the case of YapTab).

In non-trivial benchmarks, from \texttt{pg} until \texttt{peep}, which
at least in principle should reflect more accurately what one might
expect in larger applications using tabling, execution times are in
the end very competitive when comparing with XSB or YapTab. This is
probably due to the fact that the raw speed of the basic engine in
Ciao is higher than in XSB and closer to YapTab, rather than to
factors related to tabling execution, but it also implies that the
overhead of the approach to tabling used is reasonable after the
proposed optimizations in \cite{improv-tabling-padl2008}.  In this
context it should be noted that in these experiments we have used the
baseline, bytecode-based compilation and abstract machine.  Turning on
global analysis and using optimizing compilers and abstract
machines~\cite{morales04:p-to-c-padl,carro06:stream_interpreter_cases,tagschemes-ppdp08}
can further improve the speed of the SLD part of the computation.



\secend
\secend
\section{Conclusions}
\label{sec:conclusions}
\secend
\secend


We have presented an extension of the continuation call technique
which does not have the limitations of the original continuation call
approach regarding the interleaving of tabled and non-tabled predicates.
This approach has the advantage of being easier to implement and
maintain than other techniques which require non-trivial modifications
to low-level machinery.
Although there is an overhead imposed by  executing at
Prolog level, we expect the speed of the source (Prolog)
language to gradually improve by using global analysis,
optimizing compilers, and better abstract machines.  Accordingly, we
expect the performance of \texttt{CCall} to improve in the future and
thus gradually gain ground in the comparisons.

Although a non optimal tabled execution is obviously a
disadvantage, it is worth noting that, since our
implementation introduces only minimal changes in the WAM and none in the
associated Prolog compiler, the speed at which regular Prolog is
executed remains unchanged.  In addition to this, the modular design
of our approach gives better chances of making it easier to port to
other systems.  In
our case, executables which do not need tabling have very little
tabling-related code, as the data structures (for tries, etc.) are
handled by dynamic libraries loaded on demand, and only stubs are
needed in the regular engine.  The program transformation is taken
care of by a plugin for the Ciao compiler~\cite{ciaoc-entcs} (a
``package,'' in Ciao's terms) which is loaded and active only at
compile time, and which does not remain in the final executable.



\bibliographystyle{./splncs}
\bibliography{/home/clip/bibtex/clip/clip,/home/clip/bibtex/clip/general}

\end{document}